\documentstyle[12pt]{article}
\global\parskip 6pt
\begin{document}

\font\cmss=cmss10 \font\cmsss=cmss10 at 7pt
\newcommand{\room}{ & & \\ }
\newcommand{\half}{{1 \over 2}}
\newcommand{\al}{\alpha}
\newcommand{\be}{\beta}
\newcommand{\ga}{\gamma}
\newcommand{\de}{\delta}
\newcommand{\beq}{\begin{equation}}
\newcommand{\eeq}{\end{equation}}
\newcommand{\bea}{\begin{eqnarray}}
\newcommand{\eea}{\end{eqnarray}}
\newcommand{\nen}{\nonumber \\}
\renewcommand{\theequation}{\thesection.\arabic{equation}}
\newcommand{\prd}[1]{{\it Phys. Rev.} {\bf D#1}}
\newcommand{\ap}[1]{{\it Ann. Phys.} {\bf #1}}
\newcommand{\np}[1]{{\it Nucl. Phys.} {\bf B#1}}
\newcommand{\cmp}[1]{{\it Commun. Math. Phys.} {\bf #1}}
\newcommand{\ijmp}[1]{{\it Intl. J. Mod. Phys.} {\bf D#1}}
\newcommand{\ijmpa}[1]{{\it Intl. J. Mod. Phys.} {\bf A#1}}
\newcommand{\jmp}[1]{{\it J. Math. Phys.} {\bf #1}}
\newcommand{\cqg}[1]{{\it Class. Quan. Grav.} {\bf #1}}
\newcommand{\mpl}[1]{{\it Mod. Phys. Lett.} {\bf A#1}}
\newcommand{\prl}[1]{{\it Phys. Rev. Lett.} {\bf #1}}
\newcommand{\pr}[1]{{\it Phys. Rev.} {\bf #1}}
\newcommand{\pl}[1]{{\it Phys. Lett.} {\bf #1B}}
\newcommand{\rmp}[1]{{\it Rev. Mod. Phys.} {\bf #1}}
\newcommand{\prrs}[1]{{\it Proc\-.\-R.\-Soc.\-Lond.\-A. }{\bf #1}}
\newcommand{\jgp}[1]{{\it J.Geo.Phys.}{\bf #1}}
\vspace{80pt}
\begin{center}{\Huge \bf Topological Symmetries of\\
\vspace{7pt} Twisted N=2  Chiral Supergravity\\ \vspace{7pt}
in Ashtekar Formalism }\\

\vspace{30pt} \par \noindent
{P. L. Paul\footnote{Den{\'e}so{\l}in{\'e}-Treaty \# 567,
e-mail: paul@snoopy.USask.ca}}\\ \vspace{20pt}
{\sl 3208 7th St.East\\ Saskatoon, Saskatchewan \\ S7H 1B3} \\
\vspace{80pt}

\end{center}
{\bf Abstract} In this paper a topological theory of gravity is studied on a
 four-manifold using the formalism of Capovilla {\sl et al}. We show that it
is fact equivalent to Anselmi and Fre's topological gravity using the
topological symmetries. Using this
formalism gives us a new way to study topological gravity and the
intersection theory of gravitational instantons if the (3+1) decomposition
with respect to local coordinates is performed.
\newpage
\section{Introduction} \baselineskip 20pt
In recent times, Topological Field theory~\cite{TFT} has become a very popular
way to
study or compute topological invariances of oriented differentable manifolds.
 These topological invariances are computed
by counting the intersection of submanifolds of
moduli space of Instantons having zero modes. More recently
Witten found that the same invariances can be obtained
by counting the solutions to a non-linear equation. But now the
gauge group is abelian \cite{mono}. The first such theory was constructed in
the action-packed paper of
Witten \cite{Wit(TYM)}. It was constructed because of the
need for a quantum field
theoretic explanation of topological properties of some manifolds. There have
been many types of TFT that are constructed much of the work being in two
dimensions.
The topological Yang-Mills theory was constructed from twisted $N=2$
Supersymmetric
Yang-Mills theory having observables that are the Donaldson's Polynomials.\par

It has been shown in \cite{AnsFre} that twisted $N=2$ Supergravity leads
to a topological theory
of gravity(TG). Topological gravity is just the theory given in
\cite{baulsing} when the algebra is Poincar{\'e}. This paper aims
at another approach in  this direction. I will keep close contact
with their paper.\par
Another fomulation of TG has been constructed, as
a reformulation of Witten's
original form~\cite{PerTe,Wit(TG),labper,wu,karro}, namely Topological
Conformal Gravity.
This Theory uses the action which
is a analogue of TYMT, where one replaces Yang-Mills curvature
by Weyl conformal curvature tensor. In\cite{wu}, the use of Universal bundles
was used to derive the action and study the curvatures. This was done, for
Yang-Mills theory, in
\cite{baulsing} and also using Batalin-Vilkovisky  algorithm in
\cite{labper2}. With the connection between Ashtekar formulation and
Yang-Mills theory, we can hope to  use the results in TYMT to extend it to
TG in Ashtekar variables. \par
At around the same time that these TQFT were being studied, new
variable were found by Ashtekar for General Relativity~\cite{Ash,Rovrev}. It
has been used to study quantum gravity and has lead to new results.
These variables simplified the
constraints of (3+1) complex GR and have been shown to also work with
supergravity for
$N=1$ \cite{jacob} and $N=2$~\cite{KunSan}. Furthermore, since
topological gravity is exactly solvable and Ashtekar variables
simply the equations it seems to be an appropriate application.

 The problem that I would be studying in
this paper is the application of these variables to the study of
a topological theory of gravity in {\it{four dimensions}}.
The paper is ordered in the following way. First we look at the analysis of a
two-form
in general relativity. In the next section we look at the extension to
supergravity. Finally we
look at the topological twist of $N=2$ supergravity.

\setcounter{equation}{0}
\section{Chiral Form of General Relativity}
In Ashtekar formulation of general relativity , one considers
complex quantities, such as the action. Being complex does not really change
the equations of motion. The indices
$a,b...$ are tangent space indices(ie $SO(4)$ internal indices). We will
 use the two-spinor notation of Penrose and Rindler \cite{penrin} \par

 $M$ is oriented four manifold and
vanishing second Steifel-Witney class to admit spin structure. The structure
 group of the theory is
$SO(4)_C$. By analogy with real groups, and using the local isomorphism
$SU(2)^C \approx SL(2,C)$ and $SO(4) \approx
SU(2)_L \times SU(2)_R$ we state that $SU(2)_L^C \times SU(2)_R^C \approx
SO(4)_C$
\cite{gh}. These are the left and right
 handed part of Lorentz group. The fields are now sections of a
complexified vector bundle. Many of what can be said with real general
relativity can be restated in complex language.\par
 In four dimensions\cite{AHS,Don}, the $\ast$-operator maps 2-forms to 2-forms,
\beq
 *: {\bigwedge}^2 T^*M \rightarrow
{\bigwedge}^2 T^*M
\eeq which depends on the metric and has signature (++++)
and $*^2=+I$. This causes the bundle,
${\bigwedge}^2 T^*M$, to splits into self{\-}-fual and
anti{\-}-self{\-}-dual part,
\beq
{\bigwedge}^2 T^*M = {\bigwedge}^2_+ T^*M \bigoplus
{\bigwedge}^2_- T^*M .
\eeq
Therefore any two-form can be written as sum of its
self{\-}-fual and anti{\-}-self{\-}-dual part.\\
Corresponding elements are defined by the application of
the projection
operator, $P^{\pm}={\half}(1\mp*) $, to any element of the bundle. For example,
the
 projection operator acting on the
2-form made out of tetrads, $e^{a}\wedge e^{b}$ and $(su(2)_L\times
su(2)_R \rightarrow SO(4))$ give us
\bea
P^{+}(e^{a}\wedge e^{b}) &=&{\half}(e^{a}\wedge e^{b}-*e^{a}\wedge e^{b}) \nen
 &=&{\half}(e^{a}\wedge e^{b}-\epsilon^{ab}{}_{cd}e^{c}\wedge e^{d}).
\eea
notice that this is a complex quantity therefore reality conditions are
required.\par
We are on a manifold that has a metric of euclidean signature
therefore the projections created from
projection operator onto the self-dual and
anti-self-dual parts has a complex form.
The tetrad $e^{\al \be '}$,($\al$,$\be$... are the spinor indices)is the
dynamic variable rather than the metric of traditional
gravity. In four dimensional general relativity the action can be taken to be
\beq S[e^{\al \al'},\omega_{\al \be}]= \int _M e^{\al \al'}\wedge
e^{\be}_{\al'}
 \wedge R_{\al \be}
 \eeq
and $R_{\al \be}$, the curvature two-form which has the definition,
\beq R_{\al \be}=
d\omega_{\al \be}+\omega_{\al}^{\ga} \wedge \omega_{\ga \be}. \eeq
and $\omega_{\al \be}$ is $SU(2)$ connection (the self-dual part of $SO(4)$).
 The definition of Ashtekar
self-dual $SU(2)$ spin connection is
\beq
\omega^{\al \be}={\half}(\omega^{ab}-
*\omega^{ab})={\half}(\omega^{ab}-
\varepsilon^{ab}_{cd}\omega^{cd}).\eeq \par
This notion of self-dualtiy is with respect to internal indices. \par
 When self-duality is in both space-time and internal indices the Ashtekar
connection satisfies the self-dual Yang-Mills equations that come from the
variations of the action. Therefore
general relativity can be considered equivalent to self-dual Yang-Mills
theory \cite{Rovrev}.

{}From the connection we can see that
the curvature becomes
\beq
R^{\al \be}={1\over 2}(R^{ab}-*R^{ab}).\eeq which
are self-dual part of the curvature.
\par
  The tetrad can be seen to have the following combination,\beq
\Sigma^{\al \be}:=e^{\al \be'}
\wedge e^{\be}_{\al'}\label{eq:sigma}\eeq which in space-time of euclidean
signature is hermitian. It can be written as an expansion in Pauli matrices,
$\sigma^{\al \be}$,
\beq \Sigma^{\al \be}= \Sigma^k\sigma_k^{\al \be} \label{eq:pauli}\eeq
where $\Sigma^k$ is the three components as defined below.
 Equations \ref{eq:sigma} can be used as a variable \cite{israel}, but in order
for
it to have such status there must be a constraint. This constraint,
which is $\Sigma^{({\al}{\be}} \wedge \Sigma ^{\ga \de)} $, be put
 into the action with a Lagrange multiplier $\Psi_{\al \be \ga \de}$.
This makes the definition (\ref{eq:sigma}) possible. Now the action
can be written down as
\beq
S[\Sigma^{\al \be},\omega_{\al \be},\Psi_{\al \be \ga \de}]=
\int \Sigma^{\al \be}\wedge R_{\al \be}-{1 \over 2}
 \Psi_{\al \be \ga \de} \Sigma^{\al \be}
\wedge \Sigma^{\ga \de}, \eeq
 which is chiral, that is, no primes appear.
The field equations derived from this action by
variation with respect to the fields are

\beq \Sigma^{(\al \be} \wedge
 \Sigma ^{\ga \de)}=0,\label{eq:constraint}\eeq
\beq D\Sigma^{\al \be}=0,\eeq
\beq  R_{\al \be}=
\Psi_{\al \be \ga \de} \Sigma ^{\ga \de}.\eeq
As shown in \cite{Cap}, $\Psi_{\al \be \ga \de}$ is the spinor part of the
 Weyl Curvature so the curvature $R_{\al \be}$ is pure Weyl. Therefore
the Weyl curvature is self-dual.
These eq{\-}uations can be seen to be  grav{\-}itational
instantons equations \cite{kron,Cap}.
The three $ \Sigma^{\al \be}$ satisfy the relation of quaterionic
 algebra when put in the form \{ $2i \Sigma^{01}$,
 ($ \Sigma^{00}+\Sigma^{11}$),
 i($\Sigma^{00}-\Sigma^{11}$)\}.
They are the three K\"{a}hler forms of a Hyper-K\"{a}hler manifold \cite{kron},
and integrability condition, (ie $ d\Sigma^{\al \be}=0$) . This makes the
manifold hyper-K\"{a}hler(if
the term with $\Psi_{\al \be \ga \de} \Sigma ^{\ga \de}$ vanishes it is
conformally (anti) self-dual) and
 $ \Sigma^{\al \be}$ is a integrable complex
 structure on $M$. As a result, the Riemann
 curvature tensor is self-dual(in the space-time indices) and the Ricci
tensor if flat.  Any metric that has its
Riemann tensor satisfying \beq
R_{abcd}=\half \epsilon_{ablm}R^{lm}{}_{cd} \eeq is called half-flat.
 Using the Bianchi identity this implies
\beq
 R_{ab}=0 \eeq therefore gravitational instantons satisfy the
vacuum Einstein equations.
Furthermore, $R_{ab}={\half}\epsilon_{abcd}R^{cd}$ if $P^{+}\omega_{ab}=
\omega_{ab} -{\half}\epsilon_{abcd}\omega^{cd}=0$ the spin connection
is self-dual.\par
 These  gravitational instantons need not obey the self-dual Weyl
tensor \cite{ggd}.\par
 With a Cosmological constant, $\Lambda$, we get a term in the curvature
field equation,
\beq
 R_{\al \be}=\Psi_{\al \be \ga \de} \Sigma ^{\ga \de}-\frac{1}{6}
\Lambda \Sigma_{\al \be} \eeq \par

 The study of TFT of gravitational instantons
using Ashtekar variables was undertaken by Torre \cite{torre}. For
$R_{\al \be}=-\frac{1}{6} \Lambda \Sigma_{\al \be}$ and $ \Lambda > 0$ it
was shown that the moduli space of gravitational instantons is discrete, this
is the same as having its dimension to be zero.

 This form was used to study some aspects of Topological
Gravity when the term with $\Psi_{\al \be \ga \de} \Sigma ^{\ga \de}$
vanish making the manifold anti-self-dual and the cosmological constant
$\Lambda \neq 0$. \beq
 R_{\al \be}=-\frac{1}{6} \Lambda \Sigma_{\al \be}\label{eq:einstein}
 \eeq
It can be seen
to be a TYMT. This gives us
\beq
S[\omega_{\al \be}]={-6\over{\Lambda}}
\int R^{\al \be}\wedge R_{\al \be} \eeq this is the "topological charge"
and can be gauge-fixed to give Witten's Action for TYMT. And the fermionic
symmetry is just those of \cite{Wit(TYM)} and Diffeomorphism i.e. the gauge
group
is the semi-direct product of local su(2) and diffeomorphism. Combining
(\ref{eq:constraint}) and (\ref{eq:einstein}) gives us five quadratic
conditions
 that the curvature satisfies and replacing the connection by $\omega + C$
gives us a perturbation that gave the results in \cite{torre}.
 We will not go any further in this direction.
\setcounter{equation}{0}
\section{$N=2$ Chiral Supergravity in Ashtekar variables}
 In this section we will recap some results on
the Ashtekar formulation of $N=2$
Supergravity. This theory unifies gravitation with electromagnetism , and it
also
contains interaction with spin-$3\over2$ particles. See \cite{wesbag,AlvLab}
for notation and useful formulas.
 We will follow closely the paper by Kunitomo and Sano~\cite{KunSan}.
 Let ${\bf M}$ be a compact oriented manifold. In the complex
formulation, the groups $SU(2)$ are complexified. The
$N=2$ Supergravity theory has two gravitinos(the superpartner of the
graviton) forming a $SU(2)_I$ doublet and we denote it's
indices by $i,j...=1,2$. The left(right)component of the gravitino will be
denoted by
$\psi_{\al}^i$($\psi_{\al'}^i$). The indices $\al ,
\be...(\al',\be',...)=1,2$ are in the representation of
 $SU(2)_{L(R)}$. These are complex extensions
 of their real counterparts. We need to put the gravitino $\psi_{\al'}^i$
in the chiral form, which simply means redefining a field that has only
unprimed indices. We define a 2-form field as
\beq
   \chi_i^{\al}=
    e_{\al'}^{\al} \wedge \psi_i^{\al'}.
\eeq
This theory possesses a $U(1)$ symmetry giving us gauge field,
$A=A_{\mu}dx^{\mu}$,
 hence the Maxwell equation is included.\\ The chiral action for $N=2$
supergravity is \vspace{1em}
\bea
S &=& \int_{\bf M}  \Sigma^{\al \be} \wedge R_{\al \be}
 -{1 \over 2} \Psi_{\al \be \ga \de} \Sigma^{\al \be} \wedge \Sigma^{\ga \de}
 +\chi_i^{\al} \wedge
 D\psi_{\al}^i
  -\kappa_{\al \be \ga}^i \Sigma^{\al \be} \wedge \chi_i^{\ga} \nen
  & &+\phi_{\al \be}{F'} \wedge \Sigma^{\al \be}
  -{1 \over 2} \phi_{\al \be} \phi_{\ga \de} \Sigma^{\al \be}
  \wedge \Sigma^{\ga \de}+{1 \over 2}\phi_{\al \be} \chi_i^{\al}
  \wedge \chi^{i\be} \nonumber \\
& & +{1 \over 2}{F'} \wedge \psi_i^{\al} \wedge \psi_{\al}^i+
   {1 \over 8}\psi_i^{\al} \wedge \psi_{\al}^i \wedge \psi_j^{\be}
    \wedge \psi_{\be}^j,
\eea
and the auxilary field $\phi_{\al \be}$ is added to give a chiral action. $F'$
is the field
strength,
\beq
 F'=dA-{\half}\psi^i_{\al} \wedge \psi^{\al}_i. \eeq
This equation defines the supercovariant derivative.\par
Variations with respect to it's fields give us the equations of motion,,
\bea
R_{\al \be}&=& \Psi_{\al \be \ga \de}\Sigma^{\ga \de} + \kappa^i_{\al \be \ga}
\chi^\ga - \phi_{\al \be}F' +\phi_{\al \be}\phi_{\ga \de} \Sigma^{\ga \de} \nen
D\Sigma^{\al\be}&=& \chi^{(\al}_i \wedge\psi^{\be)i} \nen
D\psi^i_\al&=&\kappa^i_{\al \be \ga}\Sigma^{\be \ga}-
\phi_{\al\be}\chi^{i\be} \nen
D\chi^\al_i&=& -dA\wedge\psi^\al_i
\eea
and the contraints are, when variation is with respect to lagrange multipliera
$\Psi$, $\kappa$, and $\phi$,
\bea \Sigma^{(\al \be} \wedge \Sigma^{\ga \de)}&=&0 \nen
\Sigma^{(\al \be} \wedge \chi^{\ga)}_i &=&0 \nen
F'\wedge\Sigma^{\al \be}&=& \phi_{\ga\de}\Sigma^{\al \be}\wedge
\Sigma^{\ga\de}-
\half \chi^\al_i \wedge \chi^{i\be} \eea
The formul\ae will be useful later. A comment on the last equation, F' can
be expanded in $\Sigma$'s and arive at the condition that $\phi$ is
self-dual part of $U(1)$ curvature F.

\par
 The supersymmetries of the theory are, after combining the left and right
supersymmetries,
\bea
\delta \Sigma^{\al \be}&=&-\chi_i^{(\al}\eta^{\be)i}+
\psi_i^{(\al}\wedge \tau^{\be)i} \nonumber \\
\de \psi^i_{\al}&=&{\cal D}\eta_\al^i -\phi_{\al \be}\tau^{i \be} \nonumber \\
\de \chi_i^\al &=& F'\eta_i^\al-\phi_{\ga \de}
\Sigma^{\ga \de}\eta_i^\al +{\cal D}\tau^\al_i -\psi^\al_i \wedge \xi \nonumber
\\
\de A&=&-\psi_\al^i \eta_\al^i + \xi \nonumber \\
\de \omega_{\al \be}&=&
\kappa_{\al \be \ga}^i \tau_i^\ga- \phi_{\al \be} \xi \nen
\de \Psi_{\al \be \ga \de}&=& 0\nen
\de \kappa ^i_{\al \be \ga}&=& \Psi_{\al \be \ga \de}
 \eta^{i \de} \nen
\de \phi_{\al \be}&=& \kappa ^i_{\al \be \ga}\eta^\ga_i
. \eea
Where both the left
and right supersymmetries, $\eta^{\be}_i$, $\eta_{\be'}^i$ , respectively, have
been
combined and $C_{\al \be}$
 denote the antisymmetric
tensor(indices also for $i,j$ and $\al' \be'$). ${\cal D}$ is local Lorentz
covariant derivative \cite{ggd}. Following \cite{KunSan}, we
have the definitions
\beq \tau_i^\al \sim e_{\al'}^\al \eta_i^{\al'},\eeq
\beq \chi_i^\al =  e_{\al'}^\al \wedge \psi_i^{\al'},\eeq
and
\beq
\xi \sim  \psi_i^{\al'} \eta_{\al'}^i,
\eeq
where $\tau^\al_i$ is a fermionic spinor 1-form,and $\xi$ is a
 bosonic 1-form and the proportionality constant can be taken to be $\half$
to make the results match. Both defined by conditions,
\beq
\Sigma^{(\al \be}\wedge \tau^{\ga)}_i=0 \eeq
and
\beq
 \Sigma^{\al \be}\wedge\xi= \chi^{i(\al}\wedge \tau^{\be)}_i,\eeq
respectively. The covariant derivative is ${\cal D}$ and the cosmological
constant is zero.

\setcounter{equation}{0}
\section{Topological twist of chiral $N=2$ Supergravity}
In \cite{baulsing,labper2}, the action for TYMT was derived by BRST gauge
fixing
of a topological action. It was then shown to be to related to universal
bundles.\par
The procedure of topological twisting
$N=2$ Super Yang-Mills theory~\cite{Wit(TYM),Wit(SYM),GapOg}
can be performed in a similar way for $N=2$ Supergravity~\cite{AnsFre},
but of course  there is  a little
generalization. The action is a gauge-fixed version
of some topological action.
The gauge-fixing condition being the self-dual part of the spin
connection vanishing. This twist changes the statistic of fields, from half
integer
to integer and verse-visa.\par
 First, I will explain some of their results
using the notation of this paper. The Lorentz group is $SU(2)_L \times
SU(2)_R$, where
$SU(2)_L$ ($SU(2)_R$) is the Left-handed (Right-handed) part (these groups
being local). Also the
theory contains internal symmetry group $SU(2)_I$. Topological twist is
performed as follows: replace $SU(2)_R$ by the the diagonal subgroup
of $SU(2)_L \times SU(2)_I$. The twisted version of $SU(2)_R$ is denoted
by $SU(2)'_R$. Therefore the Lorentz Group gets replaced by $SU(2)_L
\times SU(2)'_R$, which is equivalent to internal indices the $i,j...$ being
 replaced by
$\al',\be'...$, the right-handed indices. This makes primed indices appear
all over the place.\par
 We can set as in\cite{Wit(SYM)}
\beq
 \eta^{\al \be'} = 0 ,\hspace{12pt}
\eta^{\al' \be'} = -C^{\al' \be'}\rho
 \eeq $\eta^{\al' \be'}$ is the Killing spinor used in \cite{karro} and
 from the theorem in \cite{hij} implies that the manifold is non-Kahler if
the $\eta^{\al' \be'}$ is real but it is a complex quantity in our case.
Furthermore,
 $\eta^{\al' \be'}$ is covariantly constant. \\
With these formulas at our disposal we perform the twist of the
supersymmetries and obtaining the topological symmetries. These topological
symmetries be related to the equivariant cohomology on the space of
gravitational instantons.
 We try to get formulas similar to \cite{AnsFre} but, of course, we
only get half of the results being complex version of theirs.
They also defined a shift to get the right results.
\par
First the equations above are redefined as,
\bea
 \tau^\al_{\be'} & \sim & -\rho e^\al_{\al'}\de^{\al'}_{\be'} \nonumber \\
\chi^\al_{\be'} & = & e^\al_{\al'}\wedge \psi^{\al'}_{\be'} \nonumber \\
\xi & \sim & -\rho \psi^{\al'}_{\be'}\de_{\al'}^{\be'}
\eea these are twisted version of the corresponding fields defined in the
previous section. We can keep the $\xi$ as a constant (an anticommuting
 parameter) still carrying the proportionality constant. This is
the place that fermionic symmetry and is singlet
supercharge arise and it is the component $(0,0)^{0}_{0}$ of the twisted
supersymmetry generators. This is just
translating TYMT\cite{Wit(TYM)} into supergravity, in fact, we just
generalized the procedure .
The supersymmetries now become a
 BRST-like symmetry ,
\bea
\de \Sigma^{\al \be}&=&-{\half}\rho(\psi_{\be'}^{\al}\wedge e^{\be \be'}
+\psi_{\be'}^{\be}\wedge e^{\al \be'}) \nonumber \\
\de \psi^{\be'}_{\al}&=&\rho \phi_{\al \be} e^{\be' \be}\nonumber \\
\de \chi_{\be'}^\al &=&{\cal D}\tau_{\be'}^{\al}
+\rho \psi_{\be'}^{\al}\wedge (\de_{\al'}^{\be'}\psi^{\al'}_{\be'}) \nen
\de A&=& -\rho (\de_{\al'}^{\be'}\psi^{\al'}_{\be'})\nen
\de \omega_{\al \be}&=& \kappa^{\be'}_{\al \be \de}\tau_{\be'}^{\de}
+ \rho \phi_{\al \be} (\de^{\al'}_{\be'}\psi_{\al'}^{\be'}) \nen
\de \Psi_{\al \be \ga \de}&=&0 \nen
\de \kappa^{\be'}_{\al \be \ga}&=&0 \nen
\de \phi_{\al \be}&=&0
 \eea where we can see that $\de^{\al'}_{\be'}\psi_{\al'}^{\be'}$ is already
used in \cite{AnsFre} and they called it $\tilde{\psi}$. We therefore have
another transformation \beq \de \tilde{\psi} =0 .\eeq
These symmetries can now be put into the form of BRST in suitable
redefinition of fields. Now $\de$ is the BRST-like charge $Q$. The
 first tranformation has the form $[Q,$field$]=$ghost as in
\cite{Wit(TYM)}.\par
Now put \beq
\psi^{\al \be}={\half}(
\psi_{\be'}^{\al}\wedge e^{\be \be'}+ \psi_{\be'}^{\be}\wedge e^{\al \be'})
\label{eq:topghost}\eeq
 where
$\psi^{\al \be}$ is a two-form. Also $\phi_{\al \be}$ is the spinor
form of the self-dual part of the Maxwell's curvature, $F_{\al \be}$. \par

 The term the contains the covariant derivative is proportional to the
anti-self-dual spin connection \cite{israel} minus the gravitinos,\beq
{\cal D}e^\al_{\be'}=\omega_{\be'\de'} \wedge e^{\al \de'}-\half
\psi_{\be'i}\wedge \psi^{\al i}. \eeq which came from the covariant
derivative written in two-spinor form and $\omega$ equations of motion
$D\Sigma^{\al\be}= \chi^{(\al}_i \wedge\psi^{\be)i}$ expanded in the
definition of components. This can be twisted
and the gravitinos cancel each other which is the reason for the
proportionality constant appearing in the definition of $\xi$
,i.e the term with the singlet ghost $\tilde{\psi}$ and
$\psi_{\be'}^{\al}$ vanish.

 The equations of motion \beq
{\cal D}\psi^{\be'}_\al - \kappa^{\be'}_{\al \be \ga}\Sigma^{\be \ga}+
\phi_{\al \be}\chi^{\be' \be}=0 \eeq can be used to make the identification
that $\kappa^{\be'}_{\al \be \ga} e^\ga_{\be'}$ is the self-dual part of
$\chi^{a b}$ that
appears in \cite{AnsFre}(see equation (4.4)). We write this as
${\chi'}^{\al \be}$.

\renewcommand{\theequation}{\thesection.\arabic{equation}}
The transformations become
\bea
\de \Sigma^{\al \be}&=&\rho \psi^{\al \be} \nonumber  \\
\de \psi^{\be'}_{\al}&=&-\rho \phi_{\al \be} e^{\be' \be}\nonumber \\
\de \chi_{\be'}^\al &=& \rho \omega_{\de'\be'} \wedge e^{\al\de'}\nen
\de A&=& -\rho \tilde{\psi} \nen
\de \omega_{\al \be}&=&  - \rho ({\chi'}_{\al \be}+\phi_{\al \be}\tilde{\psi}
)\eea
which can be seen to be just the unshifted BRST symmetry in \cite{AnsFre}
(see equations (4.3)) with some constants absorbed by the $\rho$'s.
By defining $\de \Phi =\rho Q \Phi$, with $\Phi$ arbitrary field in the
multiplet,
we have BRST-like operator $Q$. These equations correspond to
ref.\cite{AnsFre} if one takes the variation as the exterior derivative on some
moduli
space. For example, the first one imples
\bea
\de \Sigma^{\al \be}&= &\de (e^{\al \al'}\wedge e^{\be \be'}C_{\al' \be'})\nen
&=&(\de e^{\al \al'}C_{\al'\be'}) \wedge e^{\be \be'} + (\de e^{\be
\be'}C_{\al'\be'}
)\wedge e^{\al \al'} \eea
and putting the topological ghost (equation \ref{eq:topghost}) into the
tranformation, we get \beq
Qe^\al_{\be'}=\psi^\al_{\be'}\eeq
 It is interesting to see that the twisting procedure gave us the same
transformations as Anselmi and Fre.\par
By comparing the transformations
above to Witten's in \cite{Wit(TYM)},
$\omega_{\de'\be'}$ is the
gauge fixing condition and was realized by Anselmi and Fr{\'{e}}. They used it
in the construction of their
action for topological gravity. \par

The action for pure gravity can be seen to be of the following form \bea
{\cal L}&=&R_{\al \be}\wedge \Sigma^{\al \be} \nen
&=&2\omega^{(\al}_\de \wedge \Sigma^{\be)\de} \wedge \omega_{\al \be}-
\omega_{\al}^\de \wedge \Sigma^{\al\be} \wedge \omega_{\de \be} + d(\Sigma^{\al
\be} \wedge
\omega_{\al \be})\eea using the gravitational equations of motion for $\Sigma$.
But it is just half the story. This is true but remember that
we have taken
account the other half if the complex part of the projection onto the
 self-dual vector bundle is ignored. Giving us essentially the same theory
as Anselmi and Fr{\'e}. We may as well have rewritten the theory of Anselmi
and Fr{\'{e}} in terms of spinors.\par
The path integral can be defined formally \cite{akk} as
\bea
Z&=&\int{\cal D}\omega{\cal D}\Psi{\cal D}\Sigma \exp\{-\int\Sigma_{\al \be}
\wedge
 R^{\al \be}+\nen
& & \hspace{20pt}\half\int\Psi_{\al\be\de\ga}\Sigma^{\al \be}\wedge
\Sigma^{\de \ga}-S_{fg}-S_{FP}\}\de(tr\Psi)\nen
&=&\int{\cal D}\omega{\cal D}\Psi{\cal D}\Sigma \exp\{-\int
2\omega^{(\al}_\de \wedge \Sigma^{\be)\de} \wedge \omega_{\al \be}-\nen
& &\hspace{20pt}\omega_{\al}^\de \wedge \Sigma^{\al\be} \wedge
\omega_{\de \be}+\half\int\Psi_{\al\be\de\ga}\Sigma^{\al \be}\wedge
\Sigma^{\de \ga}-\nen
& &\hspace{20pt}S_{fg}-S_{FP}\}\de(tr\Psi)
\eea
where $S_{fg}$ and $S_{FP}$ are the gauge-fixing and Fadeev-Popov terms.
The surface terms are ignored. It remains to be seen
if such a quauntity gives a topological invariant.

\setcounter{equation}{0}
\section{Discussion}
 We have been studying the symmetries  of topological gravity from the point
of view of self-dual two forms. Which makes the Ashtekar formalism
readily available. That is achieved after the (3+1) decomposition of space-time
coordinate. \par
We can show that topological gravity comes by simply
twisting the supersymmetries of $N=2$ supergravity.  The self-dual
abelian gauge field is coupled to the gravitational
instantons making connection with \cite{mono} but things have change.
The gauge group includes the diffeomorphism group. This may be the
technique of Witten in \cite{mono} applied to the construction of
Kronheimer and Nakajima \cite{kronak}. We thus
find the topological symmetries
which need to be fixed in order that we  find the partition function.
 A few remarks about future prospect. First, in the
loop representation of the constraints, there is
  no notion of a metric. In topological field
theory, observables are metric independent\cite{TFT}. The metric
is now replaced by the densitied triad and self-dual $SU(2)$ connection.
 Therefore metric
 independence must be formulated in another way which is independence of
those variables. The area derivative
of \cite{brugpull} seems to be very convenient, also the volume
can be used but it seems that it will be more difficult. This is due to
the fact
that surfaces are better understood. Furthermore, in Donaldson's theory
we have Reimann surfaces immersed in a four-manifold and studying Dirac
operators
on these surfaces. This will be left for future work.

\section{Acknowledgments}

I would like to thank Tom Steele for his encouragement and discussions.
E. Reyes for comments and reading the first drafts. J.Szmigielski for
the tutorial on spinors. Thanks to the English River Den\'e Nation.

\end{document}